\documentclass{article}

\title{Constrained Bateman-Hillion Solutions for Hermite-Gaussian Beams }
\author{Robert Ducharme}

\begin{document}

\maketitle

\centerline{2112 Oakmeadow Pl., Bedford, TX 76021}
\centerline{E-mail: robertjducharme66@gmail.com}

\begin{abstract} 
Exact Bateman-Hillion solutions of the wave equation are applied to Hermite-Gaussian beams using a space-time constraint condition that requires the field density to fall as the inverse square of distance from the focal point of the beam at large distances from it. Following a familiar practice, the constraint is implemented in integrals through the use of a Dirac delta function. It is shown the Hermite-Gaussian functions evolve to become pure functions of angular position on the fully developed spherical phase fronts. Under the paraxial approximation it is further shown the wave equation and Schrodinger equation are interchangeable within the constraint space in correspondence to a recent paper claiming indirect evidence of Gouy phase in matter waves.
\end{abstract}

\section{Introduction}
Gouy phase is an additional phase shift that occurs in converging beams near the focal point. It has been detected in many different kinds of wave fields \cite{AES, DOM, HDM} and is of current interest in matter waves \cite{PNP}. It is usual \cite{OS} to express Gouy phase in terms of longitudinal position coordinate $x_3$ in continuous beams but the case of matter waves stands out since the time $t$ coordinate in the Schrodinger equation replaces the role of $x_3$ in the paraxial wave equation. The argument in \cite{PNP} therefore makes innovative use of a constraint space $f_P(x_3, t) = 0$ to introduce a space-time coordinate interchange symmetry that renders the differences between the Schrodinger and paraxial wave equations immaterial.

The purpose of this paper is to derive exact solutions of the wave equation for Hermite-Gaussian beams using ref. \cite{PNP} to guide the solution method. It is first noted the full wave equation contains derivatives with respect to $x_3$ like the paraxial equation and $t$ like the Schrodinger equation. A space-time symmetric solution $\Psi(x_i, t)$ is therefore sought such that $|\Psi|$ also depends on both 3-position $x_i (i=1,2,3)$ and time $t$. The time in this solution is interpreted using a constraint condition $f_E(x_i, t) = 0$. The goal being that $f_E$ should correspond to $f_P$ under the paraxial approximation.

The term constraint condition will be used throughout this paper to reference a relationship between the coordinates of space-time defining a lower dimensional constraint space. The use of constraint spaces in physics has an extensive literature for both particles and fields. Much of the work focuses on systems of interacting relativistic particles \cite{AK, CA} to manage the relative times of the particles. One common practice is the use of Dirac delta functions to impose the constraints in integrals. In the case of fields the use of the delta function constraints is also well documented in classical \cite{SR} and quantum field theory \cite{RPF} text books where the authors simply introduce the concept on intuitive grounds. Ref. \cite{PNP} does not introduce the delta function explicitly but the underlying concept is still clear.

In section 2 the known \cite{APC} idea is reviewed that if $\Psi$ is assumed to take the Bateman-Hillion form $\Phi[x_1, x_2, \frac{1}{2}(x_3+vt)-\imath a]e^{\imath k x_3 - \imath \omega t}$ where $v$ is the velocity of the beam and $a$ is a constant then the full wave equation will reduce to the form of the paraxial wave equation with $\frac{1}{2}(x_3+vt)$ replacing the traditional role of $x_3$. The converse argument is that any solution of the paraxial equation can be changed into an exact solution of the full wave equation simply by replacing every occurrence of $x_3$ in $\Phi$ with $\frac{1}{2}(x_3+vt)$.

Exact solutions for all modes of Hermite-Gaussian beams are presented in section 3. The constraint space $f_E$ for the solution is determined from the inverse square law for the intensity of the beam and it is confirmed $f_E$ reduces to $f_P$ under the paraxial approximation. In interpreting these solutions it is argued the phase fronts transition from being planar at the focal point to spherical at large distances away from it. 

In spherical polar coordinates $(r, \theta, \phi)$, the Hermite-Gaussian functions in the fully developed beam are pure functions of the inclination angle $\theta$ and azimuth angle $\phi$. The behavior of the Hermite-Gaussian patterns in relation to the phase front is therefore akin to imprinted images on a spherically shaped balloon that grows as the balloon inflates. 

A comparison of the exact Hermite-Gaussian solutions to other known solutions is presented in section 4. It is noted that another exact Gaussian mode solution exists \cite{AMT} having an identical properties under the paraxial approximation but appear dissimilar otherwise. It is further shown the exact solutions derived here corresponds to all the well known paraxial forms for the higher order modes.

It is hoped the exact Hermite-Gaussian beam solutions presented here will prove useful. It is also hope the constraint based method used to obtain them will find more general applicability for solving other partial differential equation problems.

\section{The Wave Equation for Beam Problems}
The general wave equation describes a field $\Psi(x_i, t)$ in space and time. It takes the form 
\begin{equation} \label{eq: wave_equation} 
\frac{\partial^2 \Psi}{\partial x_1^2} + \frac{\partial^2 \Psi}{\partial x_2^2} + \frac{\partial^2 \Psi}{\partial x_3^2} - \frac{1}{v^2}\frac{\partial^2 \Psi}{\partial t^2} = 0
\end{equation}
where $v$ is velocity. 

Ahead of seeking an exact solution to eq. (\ref{eq: wave_equation}) for beam problems, it is instructive to review two approximations that will together guide the way to the exact solution. One is the paraxial wave equation and the other is the Schrodinger equation.

In the paraxial approximation the solution is assumed to take the form
\begin{equation} \label{eq: paraxial_form} 
\Psi_P = \Phi_P(x_1, x_2, x_3-\imath a) \exp[\imath(k_3 x_3 -\omega t)]
\end{equation}
where $k_3$ is the component of the wave vector along $z_3$, $a$ is a constant and $\omega$ is the angular frequency. It is further assumed that $\Phi_P$ is a slowly enough varying function of $x_3$ that terms containing the second order derivatives of $\Phi_P$ with respect to $x_3$ can be neglected. This leads to the paraxial wave equation
\begin{equation} \label{eq: paraxial_wave_equation} 
\frac{\partial^2 \Phi_P}{\partial x_1^2} + \frac{\partial^2 \Phi_P}{\partial x_2^2} + 2\imath k_3 \frac{\partial \Phi_P}{\partial x_3} = 0
\end{equation}
having used $\omega =k_3v$

The Schrodinger equation for particle beams can be expressed in the form
\begin{equation} \label{eq: schrodinger_wave_equation} 
\frac{\partial^2 \Psi_S}{\partial x_1^2} + \frac{\partial^2 \Psi_S}{\partial x_2^2} + 2 \imath \frac{m}{\hbar} \frac{\partial \Psi_S}{\partial t} = 0
\end{equation}
where $m$ is the mass of each particle and $\hbar$ is Planck's constant divided by $2\pi$. It is helpful to spot $mv = k_3\hbar$. It has been argued recently \cite{PNP} that eq. (\ref{eq: schrodinger_wave_equation}) can also be written in the paraxial form
\begin{equation} \label{eq: paraxial_schrodinger_equation} 
\frac{\partial^2 \Psi_S}{\partial x_1^2} + \frac{\partial^2 \Psi_S}{\partial x_2^2} + 2\imath k_3 \frac{\partial \Psi_S}{\partial x_3} = 0
\end{equation}
for solutions restricted to the constraint space
\begin{equation} \label{eq: constraint_simple} 
f_P(x_3, t) = x_3 - vt = 0
\end{equation}
This finding is part of an argument used to imply that Gouy phase should exist in matter waves the same as it does in other kinds of waves.  

The foregoing argument contains two interesting ideas that will next be elevated to the status of guiding principles for the purpose of finding a more general solution of the wave equation (\ref{eq: wave_equation}) for exact forms of $\Psi$. One is that $x_3$ and $vt$ are interchangeable in the paraxial approximation through eq. (\ref{eq: constraint_simple}). There is therefore merit in prioritizing potential trial solutions for $\Psi$ that include this interchangeability symmetry in the $\Phi$ component of the solution. The other is the restriction to the constraint space (\ref{eq: constraint_simple}). The requisite correspondence principle in this case is to seek solutions for $\Psi$ in a constraint space of the form $f_E(x_i, t) = 0$ such that $f_E(x_i, t)$ reduces to $f_P(x_3, t)$ under the paraxial approximation.

The concept of finding trial solutions for $\Psi$ that preserve their form under the interchange of $x_3$ and $vt$ in the $\Phi$ component of the wave function suggests replacing all occurrences of $x_3$ in eq. (\ref{eq: paraxial_form}) with $\frac{1}{2}(x_3+vt)$ to give
\begin{equation} \label{eq: exact_form} 
\Psi = \Phi \left[x_1, x_2, \frac{1}{2} (x_3+vt)-\imath a \right] \exp[\imath(k_3 x_3 -\omega t)]
\end{equation}
As a solution of this form must be assumed to exist in a constraint space, the meaning of the solution will not become clear until the form of $f_E(x_i, t)$ is determined. One point of possible confusion here is that $\Psi$ contains a planar phase factor even though it has been made clear earlier that beams have spherical phase fronts. The resolution to this apparent contradiction is that $\Psi$ can have planar phase fronts in mathematical 4-space but it must have spherical phase fronts in the physical constraint space where it will be interpreted.

Eq. (\ref{eq: exact_form}) generates the second order derivatives
\begin{equation} \label{eq: second_deriv_z} 
\frac{\partial^2 \Psi}{\partial x_3^2} = \exp[\imath(k_3 x_3 -\omega t)]\left( \frac{\partial^2 \Phi}{\partial x_3^2} + 2\imath k_3 \frac{\partial \Phi}{\partial x_3} - k_3^{2} \Phi \right)
\end{equation}
\begin{equation} \label{eq: second_deriv_t} 
\frac{\partial^2 \Psi}{\partial t^2} = \exp[\imath(k_3 x_3 -\omega t)]\left( \frac{\partial^2 \Phi}{\partial t^2} - 2\imath \omega \frac{\partial \Phi}{\partial t} - \omega^{ 2} \Phi \right)
\end{equation}
The interchangeability of $x_3$ and $vt$ in the solution (\ref{eq: exact_form}) implies
\begin{equation} \label{eq: first_symmetry} 
\frac{\partial \Phi}{\partial x_3} = \frac{1}{v} \frac{\partial \Phi}{\partial t} 
\end{equation}
\begin{equation} \label{eq: second_symmetry} 
\frac{\partial^2 \Phi}{\partial x_3^2} = \frac{1}{v^2} \frac{\partial^2 \Phi}{\partial t^2} 
\end{equation}
Inserting eqs. (\ref{eq: second_deriv_z}) and (\ref{eq: second_deriv_t}) into the wave equation (\ref{eq: wave_equation}) and simplifying the resulting expression using eqs. (\ref{eq: first_symmetry}) and (\ref{eq: second_symmetry}) gives
\begin{equation} \label{eq: wave_equation_subset} 
\frac{\partial^2 \Phi}{\partial x_1^2} + \frac{\partial^2 \Phi}{\partial x_2^2} + 2\imath k_3 \frac{\partial \Phi}{\partial s} = 0
\end{equation}
having put
\begin{equation} \label{eq: s_param} 
s = \frac{1}{2}(x+vt)
\end{equation}
It is thus concluded eq. (\ref{eq: wave_equation}) has a class of solutions of the form (\ref{eq: exact_form}) that also satisfy eq. (\ref{eq: wave_equation_subset}).

\section{Exact Hermite-Gaussian Solutions}
It is interesting that eq. (\ref{eq: wave_equation_subset}) has the same mathematical form as the paraxial wave equation (\ref{eq: paraxial_wave_equation}) with $s$ substituting for $x_3$. This implies that any solution of the paraxial equation can be turned into an exact solution of the full wave equation eq. (\ref{eq: wave_equation}) using just the $x_3 \rightarrow s$ replacement providing a meaningful constraint space can be found to interpret the solution.

The wave equation (\ref{eq: wave_equation_subset}) can be solved for a complete orthonormal basis set of Hermite-Gaussian functions \cite{AES}. These take the form
\begin{equation} \label{eq: hermite_gauss_solution} 
\Phi_{mn} = \frac{C_{mn} w_0}{w(s)}H_m\left( \frac{\sqrt{2}x_1}{w(s)}\right) H_n\left( \frac{\sqrt{2}x_2}{w(s)}\right)\exp \left[ \frac{\imath k \rho^2}{2(s-\imath L_R)} - \imath g_{mn}(s) \right]
\end{equation}
where
\begin{equation} \label{eq: spot_radius} 
w(s) = w_0\sqrt{1+\left( \frac{s}{L_R} \right)^2}
\end{equation}
is the radius of the laser spot, $s-\imath L_R$ is the complex beam parameter and
\begin{equation} \label{eq: gouy_phase} 
g_{mn}(s) = (1+m+n)\arctan \frac{s}{L_R}
\end{equation}
is the Gouy phase. In this, $w_0=w(0)$ is the radius of the beam waist at $s=0$, $L_R = \frac{1}{2}kw_0^2$ is the Rayleigh range and $H_m$ and $H_n$ are Hermite polynomials where $m$ and $n$ are positive integers. The normalizing constant $C_{mn}$ is chosen to give
\begin{equation} \label{eq: scalar_orthonormal} 
\int_{-\infty}^{+\infty} \Phi_{pq}^* \Phi_{mn} \delta[f_E(x_i, t)]d^3xdt = \delta_{mp}\delta_{nq}
\end{equation}
where $\delta_{mp}$ is the Kronecker delta and $\delta[f(x_i, t)]$ is the Dirac delta function.

It is readily verified through direct substitution that eqs. (\ref{eq: exact_form}) and (\ref{eq: hermite_gauss_solution}) satisfy eq. (\ref{eq: wave_equation}). The next step is to determine the form of the $f_E(x_i, t)=0$ constraint space so the solution can be interpreted. For this purpose, it will be sufficient to require the field density $|\Psi_{mn}|^2$ to take the inverse square law form
\begin{equation} \label{eq: inverse_square_law} 
D_{mn} = \int_{-\infty}^{+\infty} |\Psi_{mn}(r, \theta, \phi, t)|^2\delta[f_E(x_i, t)]dt = \frac{F_{mn}(\theta, \phi)}{r^2}
\end{equation}
for $r (=\sqrt{x_1^2+x_2^2+x_3^2}) \gg L_R$ consistent with the idea the beam develops over large distances to have spherically symmetric phase fronts. 

The explicit form of $D_{mn}$ can be calculated from eqs. (\ref{eq: exact_form}) and (\ref{eq: hermite_gauss_solution}) to give
\begin{equation} \label{eq: density_function} 
\frac{C_{mn}^2 w_0^2}{w^2(s)}H_m^2\left( \frac{\sqrt{2}r \sin \theta \cos \phi}{w(s)}\right) H_n^2\left( \frac{\sqrt{2}r \sin \theta \sin \phi}{w(s)}\right)\exp \left[ - \frac{2 r^2 \sin^2 \theta}{w^2(s)} \right]
\end{equation}
approximating to
\begin{equation} \label{eq: density_function_approx} 
\frac{C_{mn}^2}{(s/L_R)^2}H_m^2\left( \frac{\sqrt{2} r \sin \theta \cos \phi}{ w_0 (s/L_R)}\right) H_n^2\left( \frac{\sqrt{2} r \sin \theta \sin \phi}{ w_0 (s/L_R)}\right)\exp \left[ - \frac{2 r^2 \sin^2 \theta}{w_0^2 (s/L_R)^2} \right]
\end{equation}
for large values of $s(\gg L_R)$. It follows therefore that $|\Psi|^2 \propto \frac{1}{r^2}$ for $r \gg L_R$ if and only if
\begin{equation} \label{eq: constraint_hermite_gauss_1} 
f_E(x_i, t)= r - s = 0
\end{equation}
or equivalently
\begin{equation} \label{eq: constraint_hermite_gauss_2} 
f_E(x_i, t)= r - \frac{1}{2}(x_3+vt) = 0
\end{equation}
In particular, inserting eq. (\ref{eq: constraint_hermite_gauss_1}) into (\ref{eq: density_function_approx}) gives the explicit form of  $F_{mn}(\theta, \phi)$ to be 
\begin{equation} \label{eq: density_function_approx_r} 
C_{mn}^2 L_R^2 H_m^2\left( \frac{\sqrt{2} \sin \theta \cos \phi}{ (w_0/L_R)}\right) H_n^2\left( \frac{\sqrt{2} \sin \theta \sin \phi}{ (w_0/L_R)}\right)\exp \left[ - \frac{2 \sin^2 \theta}{(w_0/L_R)^2} \right]
\end{equation}
Eq. (\ref{eq: constraint_hermite_gauss_2}) will therefore be taken as the constraint condition defining the constraint space for $\Psi$. It is further interest that $f_E(x_i, t)$ does correspond to $f_P(x_3, t)$ in the paraxial approximation as expected since $x_3 \simeq r$ in this limit.

Overall, it is understood the phase fronts transition from being planar near the focal point of the beam to spherical as the beam becomes fully developed at large distances $r \gg L_R$ from the focal point. It is also been shown that the Hermite-Gaussian component of the beam takes the form $F_{mn}(\theta, \phi)$ in the fully developed beam indicating that it is purely a function of angular position on the spherical fronts.

\section{Comparison to Other Solutions}
The solution to the paraxial wave equation (\ref{eq: paraxial_wave_equation}) for Hermite-Gaussian beams is readily recovered from the exact solution (\ref{eq: hermite_gauss_solution}) through a simple application of the paraxial constraint condition (\ref{eq: constraint_simple}). This gives
\begin{equation} \label{eq: paraxial_hermite_gauss_solution} 
\Phi_{mn} = \frac{C_{mn} w_0}{w(x_3)}H_m\left( \frac{\sqrt{2}x_1}{w(x_3)}\right) H_n\left( \frac{\sqrt{2}x_2}{w(x_3)}\right)\exp \left[ \frac{\imath k \rho^2}{2(x_3-\imath L_R)} - \imath g_{mn}(x_3) \right]
\end{equation}
where
\begin{equation} \label{eq: paraxial_spot_radius} 
w(x_3) = w_0\sqrt{1+\left( \frac{x_3}{L_R} \right)^2}
\end{equation}
is the radius of the laser spot, $x_3-\imath L_R$ is the complex beam parameter and
\begin{equation} \label{eq: paraxial_gouy_phase} 
g_{mn}(x_3) = (1+m+n)\arctan \frac{x_3}{L_R}
\end{equation}
is the Gouy phase. The complete solution is just $\Psi_{mn} = \Phi_{mn}e^{\imath k x_3 - \imath \omega t}$.

The wave equation (\ref{eq: wave_equation}) has another known \cite{APK} exact solution for Gaussian mode beams different from the one derived here in eqs. (\ref{eq: exact_form}) and (\ref{eq: hermite_gauss_solution}):
\begin{equation} \label{eq: gaussian_solution} 
\Psi_{00} = \frac{C_{00}L_R}{L_R+\imath \frac{1}{2}(x_3+vt)} \exp \left( \frac{\imath k \rho^2}{x_3+vt-\imath 2L_R} \right) \exp[\imath(k x_3 -\omega t)]
\end{equation}
This other solution takes the form
\begin{equation} \label{eq: other_exact_solution} 
\Psi_{00}^\prime = \frac{D_{00}L_R}{R}\exp[\imath(kR-\omega t)]
\end{equation}
where $R =\sqrt{x_1^2+x_2^2+(x_3-\imath L_R)^2}$. The relationship of this result to a Gaussian mode beam can be made clear under the paraxial approximation where
\begin{equation} \label{eq: paraxial_R} 
R \simeq x_3 - \imath L_R + \frac{\rho^2}{2(x_3 - \imath L_R)}
\end{equation}
Inserting this expression into eq. (\ref{eq: other_exact_solution}) gives
\begin{equation} \label{eq: paraxial_gaussian_solution} 
\Psi_{00}^\prime \simeq \frac{C_{00}L_R}{L_R+\imath x_3} \exp \left[ \frac{\imath k \rho^2}{2(x_3-\imath L_R)} \right] \exp[\imath(k x_3 -\omega t)]
\end{equation}
having neglected the second order term in the denominator and set $C_{00} = \imath\exp(kL_R)D_{00}$. It can be seen therefore that eqs. (\ref{eq: gaussian_solution}) and (\ref{eq: other_exact_solution}) both correspond to the same paraxial form  (\ref{eq: paraxial_gaussian_solution}) for nearly parallel beams but appear to belong to different solution classes. It will therefore take more work to properly understand the nature of the relationship between these two exact solutions of the wave equation.

\section{Summary}
A recent paper \cite{PNP} on Gouy phase in matter waves has indicated the existence of a symmetry that can be imposed through a constraint space to render the Schrodinger and paraxial wave equations interchangeable. In this paper, constrained Bateman-Hillion solutions have been obtained for Hermite-Gaussian beams that incorporate this symmetry and therefore also require interpretation using a constraint space. It has been found the form of this constraint space is readily determined through the simple requirement that the intensity of the beam varies as the inverse square of distance from the focal point of the beam at large distances from it. The solutions indicate the phase fronts transition from a planar form near to the focal point to a spherical form in the fully developed beam. It has also been found the Hermite-Gaussian functions evolve to become pure functions angular position on the spherical phase fronts.

\end{document}